\documentclass{aa}
\usepackage{graphicx}
\usepackage{times}
\begin{document}
\thesaurus{05                       
          (10.15.2 (NGC\,1960, NGC\,2194);  
           05.01.1;                 
           08.11.1;                 
           08.08.1;                 
           08.12.3)                 
           }

\title{Photometric and kinematic studies of open star clusters.\\
II. NGC\,1960 (M 36) and NGC\,2194\thanks{partly based on data observed at the
  German-Spanish Astronomical Centre, Calar Alto, operated by the
  Max-Planck-Institute for Astronomy, Heidelberg, jointly with the Spanish
  National Commission for Astronomy}}

\author{J\"org Sanner\inst{1}, Martin Altmann\inst{1}, Jens Brunzendorf\inst{2}, Michael Geffert\inst{1}}

\institute{Sternwarte der Universit\"at Bonn, Auf dem H\"ugel 71, 
           D--53121 Bonn, F.R. Germany \and
           Th\"uringer Landessternwarte Tautenburg, Sternwarte 5,
           D--07778 Tautenburg, F.R. Germany}

\offprints{J\"org Sanner, jsanner@astro.uni-bonn.de}

\date{Received 7 December 1999, accepted 14 March 2000}

\maketitle

\markboth{J. Sanner et al.: Photometric and kinematic studies of open star
  clusters. II}{J. Sanner et al.: Photometric and kinematic studies of open
  star clusters. II}

\begin{abstract}
We present CCD photometry and proper motion studies of the two open star
clusters NGC\,1960 (M\,36) and NGC 2194. Fitting isochrones to the colour
magnitude diagrams, for NGC 1960 we found an age of $t=16$ Myr and a distance
of roughly $d=1300$ pc and for NGC 2194 $t=550$ Myr and $d=2900$ pc,
respectively. We combined membership determination by proper motions and
statistical field star subtraction to derive the initial mass function of the
clusters and found slopes of $\Gamma=-1.23 \pm 0.17$ for NGC\,1960 and
$\Gamma=-1.33 \pm 0.29$ for NGC\,2194. Compared to other IMF studies of the
intermediate mass range, these values indicate shallow mass functions.


\end{abstract}

\section{Introduction}

The shape of the initial mass function (IMF) is an important parameter to
understand the fragmentation of molecular clouds and therefore the formation
and development of stellar systems. Besides studies of the Solar neighbourhood
(Salpeter \cite{salpeter}, Tsujimoto et al. \cite{tsuji}), work on star
clusters plays a major role (Scalo \cite{scalo1}) in this field, as age,
metallicity, and distance of all stars of a star cluster can generally be
assumed to be equal.

Restricted to certain mass intervals, the IMF can be described by a power law
in the form
\begin{equation}
\mbox{d} \log N(m) \sim m^\Gamma \mbox{d} \log m.
\end{equation}
In this
notation the ``classical'' value found by Salpeter (\cite{salpeter}) for the
Solar neighbourhood is $\Gamma=-1.35$. Average values for $\Gamma$ from more
recent studies, mostly of star clusters, can be found, e.g., in Scalo
(\cite{scalo2}):
\begin{eqnarray}
\Gamma=-1.3 \pm 0.5 & \mbox{\ for } & m > 10 M_\odot, \nonumber \\
\Gamma=-1.7 \pm 0.5 & \mbox{\ for } & 1 M_\odot < m < 10 M_\odot, \mbox{\ and}\\
\Gamma=-0.2 \pm 0.3 & \mbox{\ for } & m < 1 M_\odot, \nonumber
\end{eqnarray}
where the ``$\pm$'' values refer to a rough range of the slopes derived for the
corresponding mass intervals, caused by empirical uncertainties or probable
real IMF variations.

Knowledge of membership is essential to derive the IMF especially of open star
clusters, where the contamination of the data with field stars presents a
major challenge. Two methods for field star subtraction are in use nowadays:
separating field and cluster stars by means of membership probabilities from
stellar proper motions on one hand, statistical field star subtraction on the
other hand. Our work combines these two methods: The proper motions are
investigated for the bright stars of the clusters, down to the completeness
limit of the photographic plates used, whereas the fainter cluster members are
selected with statistical considerations.

From the cleaned data we derive the luminosity and mass functions of the
clusters. Including the proper motions, we expect to receive a more reliable
IMF, since the small number of bright stars in open clusters would lead to
higher uncertainties, if only statistical field star subtraction were
applied.

This is the second part of a series of studies of open star clusters, following
Sanner et al. (\cite{n0581paper}). Here we present data on two clusters of the
northern hemisphere, NGC\,1960 (M\,36) and NGC\,2194.

NGC\,1960 (M\,36) is located at $\alpha_{2000}=5^{\rm h}
36^{\rm m}6^{\rm s}$, $\delta_{2000}=+34^\circ 8 \arcmin$ and has a
diameter of $d=10 \arcmin$ according to the Lyng{\aa} (\cite{lynga})
catalogue. Morphologically, NGC\,1960 is dominated by a number of bright
($V \ga 11 \mbox{\ mag}$) stars, whereas the total stellar density is only
marginally enhanced compared to the surrounding field. The cluster has
not yet been studied by means of CCD photometry. Photographic photometry was
published by Barkhatova et al. (\cite{barkhatova}), photoelectric
photometry of 50 stars in the region of the cluster by Johnson \& Morgan
(\cite{johnsmorg}). The most recent proper motion studies are from Meurers
(\cite{meurers}) and Chian \& Zhu (\cite{chianzhu}). As their epoch
differences between first and second epoch plates (36 and 51 years,
respectively) are smaller than ours and today's measuring techniques can be
assumed to be more precise we are confident to gain more reliable results.

Tarrab (\cite{tarrab}) published an IMF study of 75 open star clusters,
among them NGC\,1960, and found an exteme value for the slope of (in our
notation) $\Gamma=-0.24 \pm 0.05$ for this object. Her work includes only 25
stars in the mass range $3.5 M_\odot \la m \la 9 M_\odot$, so that a more
detailed study covering more members and reaching towards smaller masses is
necessary.

For NGC\,2194 (located at $\alpha_{2000}=6^{\rm h} 13^{\rm m}48^{\rm s}$,
$\delta_{2000}=+12^\circ 48 \arcmin$, diameter $d=9 \arcmin$), our work is the
first proper motion study according to van Leeuwen (\cite{vanleeuwen}).
The {\it RGU} photographic photometry of del\,Rio (\cite{delrio}) is the most
recent publication on NGC\,2194 including photometric work.

The cluster is easily detectable as it contains numerous intermediate
magnitude ($13 \mbox{\ mag} \la V \la 15 \mbox{\ mag}$) stars, although bright
stars $V \la 10 \mbox{\ mag}$ are lacking.

In Sect. \ref{cadata}, we present the data used for our studies and the basic
steps of data reduction and analysis. Sects. \ref{n1960disc} and
\ref{n2194disc} include the proper motion studies, an analysis of the colour
magnitude diagrams (CMDs), and determination of the IMF of the clusters. We
conclude with a summary and discussion in Sect. \ref{caconcl}.

\section{The data and data reduction}

\label{cadata}

\subsection{Photometry} \label{caphot}

\begin{figure}
\centerline{
\includegraphics[width=\hsize]{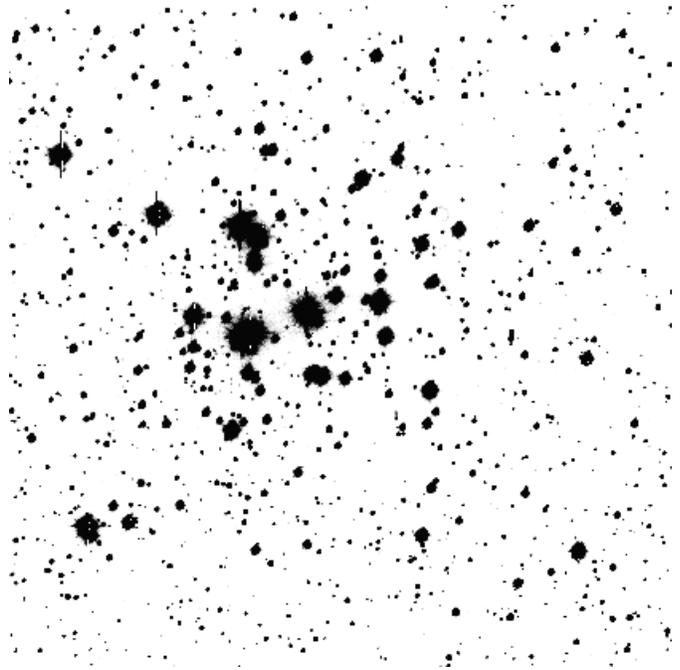}
}
\caption[]{\label{n1960bild} $600$ s $V$ CCD image of
  NGC\,1960. The field of view is approximately $14 \arcmin \times 14 \arcmin$,
  north is up, east to the left}
\end{figure}

\begin{figure}
\centerline{
\includegraphics[width=\hsize]{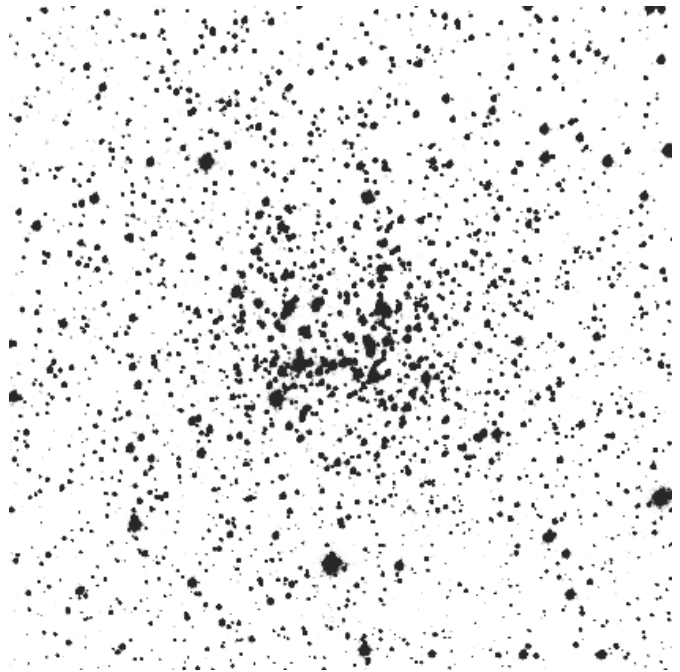}
}
\caption[]{\label{n2194bild} $600$ s $V$ CCD image of NGC\,2194. As in
  Fig. \ref{n1960bild}, the field of view is approximately $14 \arcmin \times
  14 \arcmin$ with north up and east to the left}
\end{figure}

CCD images of both clusters were taken with the 1.23\,m telescope at Calar Alto
Observatory on October 15, 1998, in photometric conditions. The seeing was of
the order of $3 \arcsec$. The telescope was equipped with the $1024 \times
1024$ pix CCD chip TEK 7\_12 with a pixel size of $24 \mbox{\ } \mu \mbox{m}
\times 24 \mbox{\ } \mu \mbox{m}$ and the WWFPP focal reducing system (Reif et
al. \cite{wwfpp}). This leads to a resolution of $1.0 \arcsec \mbox{\ pix}
^{-1}$ and a field of view of $17 \arcmin \times 17 \arcmin$. Both clusters
were observed in Johnson $B$ and $V$ filters, the exposure times were 1\,s,
10\,s, and 600\,s in $V$, and 2\,s, 20\,s, and 900\,s in
$B$. Figs. \ref{n1960bild} and \ref{n2194bild} show CCD images of both
clusters.

The data were reduced with the DAOPHOT II software (Stetson \cite{daophot})
running under IRAF. From the resulting files, we deleted all objects showing
too high photometric errors as well as sharpness and $\chi$ values. The limits
were chosen individually for each image, typical values are $0.03 \mbox{\
  mag}$ to $0.05 \mbox{\ mag}$ for the magnitudes, $\pm 0.5$ to $1$ for
sharpness, and $2$ to $4$ for $\chi$.

Resulting photometric errors of the calibrated magnitudes in different $V$
ranges valid for both clusters as given by the PSF fitting routine are given
in Table \ref{caphoterrors}.

\begin{table}
\caption[]{\label{caphoterrors}Typical photometric errors for stars in
  different magnitude ranges}
\begin{tabular}{r@{$V$}lcc}
\hline
\multicolumn{2}{c}{magnitude} & $\Delta V$ & $\Delta (B-V)$ \\
\multicolumn{2}{c}{range} & $[\mbox{mag}]$ & $[\mbox{mag}]$\\
\hline
         & $< 12 \mbox{\ mag}$ & $0.01$ & $0.01$ \\
 $12 \mbox{\ mag} <$  & $< 16 \mbox{\ mag}$ & $0.02$ & $0.04$ \\
 $16 \mbox{\ mag} <$  &        & $0.04$ & $0.08$ \\
\hline
\end{tabular}
\end{table}

The data were calibrated using 44 additional observations of a total of 27
Landolt (\cite{landolt}) standard stars. After correcting the instrumental
magnitudes for atmospheric extinction and to exposure times of 1\,s, we used
the following equations for transformation from instrumental to apparent
magnitudes:
\begin{eqnarray}
v-V &=& z_V - c_V \cdot (B-V)\\
(b-v)-(B-V) &=&  z_{B-V} - c_{B-V} \cdot (B-V)
\end{eqnarray}
where capital letters represent apparent and lower case letters (corrected as
described above) instrumental magnitudes. The extinction coefficients $k_V$
and $k_{B-V}$, zero points $z_V$ and $z_{B-V}$ as well as the colour terms
$c_V$ and $c_{B-V}$ were determined with the IRAF routine {\tt fitparams} as:
\begin{eqnarray}
k_V=0.14 \pm 0.02, & & k_{B-V}=0.19 \pm 0.03 \nonumber\\
z_V=2.52 \pm 0.04, & & z_{B-V}=0.88 \pm 0.04\\
c_V=0.09 \pm 0.01, & & c_{B-V}=0.19 \pm 0.01. \nonumber
\end{eqnarray}
We checked the quality of these parameters by reproducing the apparent
magnitudes of the standard stars from the measurements. The standard
deviations derived were $\sigma_V=0.02 \mbox{\ mag}$ and $\sigma_{B-V}=0.06
\mbox{\ mag}$.

Johnson \& Morgan (\cite{johnsmorg}) published photoelectic photometry of
50 stars in the region of NGC\,1960. Their results coincide with ours with a
standard deviation of approx. $0.03 \mbox{\ mag}$ in $V$ and $0.02 \mbox{\
mag}$ in $B-V$, respectively. There is only one exception, star 110 (Boden's
(\cite{boden}) star No. 46) for which we found $V=14.25 \mbox{\ mag}$,
$B-V=0.66 \mbox{\ mag}$, which differs by $\Delta V \approx 2 \mbox{\
mag}$ and $\Delta B-V \approx 0.3 \mbox{\ mag}$ from the value of Johnson \&
Morgan (\cite{johnsmorg}). In their photographic photometry, Barkhatova et
al. (\cite{barkhatova}) found values for this star which coincide with
ours. We therefore assume the difference most likely to be caused by a
mis-identification of this star by Johnson \& Morgan (\cite{johnsmorg}).

\begin{figure}
\centerline{
\includegraphics[width=\hsize]{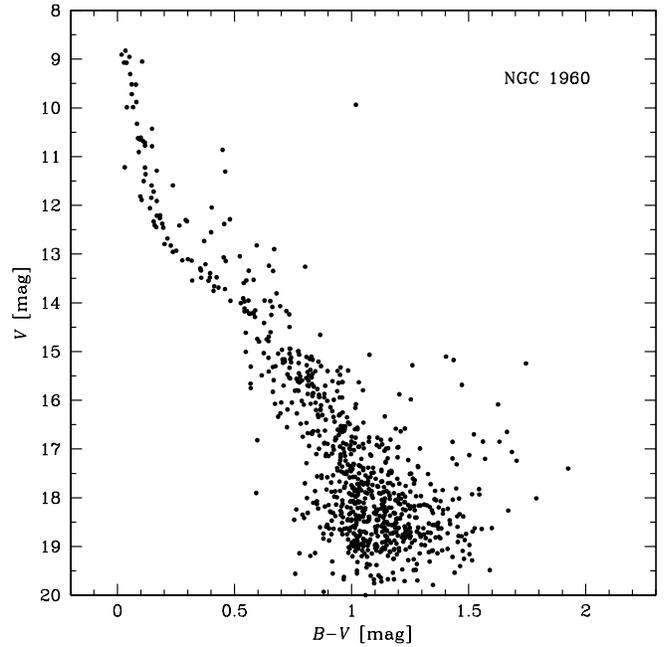}
}
\caption[]{\label{n1960cmd} Colour magnitude diagram of NGC\,1960 and the
  surrounding field. This diagram contains all stars for which both $B$ and
  $V$ magnitudes were determined. Stars with too high photometric errors were
  excluded beforehand. This CMD is still contaminated with field stars. For a
  CMD which is field star corrected see Fig. \ref{n1960cmdhaufen}}
\end{figure}

\begin{figure}
\centerline{
\includegraphics[width=\hsize]{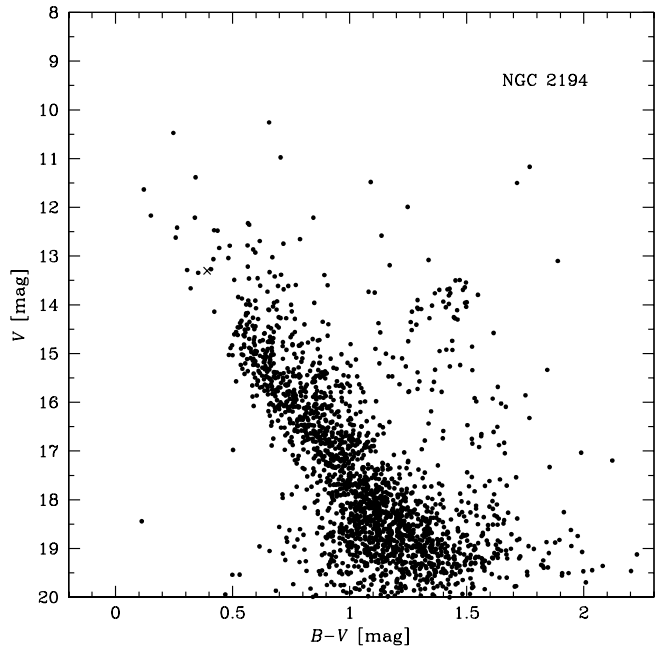}
}
\caption[]{\label{n2194cmd} Colour magnitude diagram of all stars in the field
  of NGC 2194. For further remarks see Fig. \ref{n1960cmd}. The star marked
  with a cross is claimed to be a blue straggler by Ahumada \& Lapasset
  (\cite{bluestrag}). This statement is discussed in
  Sect. \ref{n2194cmdsect}. Fig. \ref{n2194cmdhaufen} shows the field star
  corrected CMD of NGC\,2194}
\end{figure}

All stars for which $B$ and $V$ magnitudes could be determined are listed in
Tables \ref{n1960cdsphot} (NGC\,1960, 864 stars) and \ref{n2194cdsphot}
(NGC 2194, 2120 stars), respectively. We derived the CMDs of the two clusters
which are shown in Figs. \ref{n1960cmd} and \ref{n2194cmd}. A detailed
discussion of the diagrams is given in Sects. \ref{n1960disc} and
\ref{n2194disc}.

\begin{table}
\caption[]{\label{n1960cdsphot} List of the photometric data of all stars
  measured in the CCD field of NGC\,1960. For cross-identification, the
  star numbers of Boden (\cite{boden}) are given, too. Only the ten brightest
  stars for which both photometry and proper motions were determined are
  listed here, the complete table is available online at the CDS archive}
\begin{tabular}{rrrrrr}
\hline
\multicolumn{1}{c}{No.} & \multicolumn{1}{c}{Boden} & \multicolumn{1}{c}{$x$} & \multicolumn{1}{c}{$y$} &
\multicolumn{1}{c}{$V$} & \multicolumn{1}{c}{$B-V$} \\
& \multicolumn{1}{c}{No.} &  &  & \multicolumn{1}{c}{[mag]} & \multicolumn{1}{c}{[mag]} \\
\hline
$ 1$ & $   $ & $432.729$ & $550.012$ & $8.83$ & $+0.03$\\
$ 2$ & $ 23$ & $349.018$ & $441.873$ & $8.91$ & $+0.02$\\
$ 3$ & $138$ & $157.530$ & $819.617$ & $8.95$ & $+0.05$\\
$ 4$ & $101$ & $124.550$ & $353.711$ & $9.05$ & $+0.11$\\
$ 5$ & $ 61$ & $245.181$ & $427.961$ & $9.07$ & $+0.03$\\
$ 8$ & $ 27$ & $291.005$ & $555.456$ & $9.52$ & $+0.06$\\
$ 9$ & $ 48$ & $524.133$ & $537.383$ & $9.52$ & $+0.08$\\
$10$ & $ 21$ & $372.126$ & $456.779$ & $9.72$ & $+0.06$\\
$11$ & $ 38$ & $339.012$ & $698.830$ & $9.88$ & $+0.08$\\
$12$ & $184$ & $774.058$ & $850.419$ & $9.94$ & $+1.02$\\
\multicolumn{1}{c}{$\vdots$} & \multicolumn{1}{c}{$\vdots$} &
\multicolumn{1}{c}{$\vdots$} & \multicolumn{1}{c}{$\vdots$} &
\multicolumn{1}{c}{$\vdots$} & \multicolumn{1}{c}{$\vdots$} \\
\hline
\end{tabular}
\end{table}

\begin{table}
\caption[]{\label{n2194cdsphot} List of the photometric data of all stars
  measured in the CCD field of NGC\,2194. As for Table \ref{n1960cdsphot},
  only the ten brightest stars for which we derived photometric data and
  proper motions are mentioned here. As a cross reference, the numbers
  from del\,Rio (\cite{delrio}) are added. The complete table is available at
  the CDS archive}
\begin{tabular}{rcrrrr}
\hline
\multicolumn{1}{c}{No.} & del\,Rio & \multicolumn{1}{c}{$x$} & \multicolumn{1}{c}{$y$} &
\multicolumn{1}{c}{$V$} & \multicolumn{1}{c}{$B-V$} \\
& \multicolumn{1}{c}{No.} &  &  & \multicolumn{1}{c}{[mag]} & \multicolumn{1}{c}{[mag]} \\
\hline
$ 1$ & $   $ & $ 591.957$ & $790.526$ & $10.26$ & $+0.66$\\
$ 2$ & $   $ & $1010.094$ & $706.577$ & $10.47$ & $+0.25$\\
$ 3$ & $   $ & $ 434.371$ & $282.158$ & $10.97$ & $+0.71$\\
$ 4$ & $ 39$ & $ 522.864$ & $581.859$ & $11.17$ & $+1.77$\\
$ 5$ & $   $ & $ 978.803$ & $947.148$ & $11.38$ & $+0.34$\\
$ 6$ & $   $ & $ 344.500$ & $740.752$ & $11.48$ & $+1.09$\\
$ 7$ & $ 49$ & $ 552.482$ & $537.831$ & $11.50$ & $+1.72$\\
$ 8$ & $   $ & $ 634.918$ & $899.808$ & $11.63$ & $+0.12$\\
$10$ & $ 14$ & $ 646.659$ & $554.135$ & $12.16$ & $+2.40$\\
$11$ & $   $ & $ 867.565$ & $756.487$ & $12.17$ & $+0.15$\\
\multicolumn{1}{c}{$\vdots$} & \multicolumn{1}{c}{$\vdots$} &
\multicolumn{1}{c}{$\vdots$} & \multicolumn{1}{c}{$\vdots$} &
\multicolumn{1}{c}{$\vdots$} & \multicolumn{1}{c}{$\vdots$}\\
\hline
\end{tabular}
\end{table}

\subsection{Actual cluster sizes} \label{cagroesse}

Mass segregation might lead to a larger ``true'' cluster size than stated,
e.g., in the Lyng{\aa} (\cite{lynga}) catalogue: While the high mass stars are
concentrated within the inner part of the cluster, the lower mass stars might
form a corona which can reach as far out as the tidal radius of the cluster
(see, e.g., the recent work of Raboud \& Mermilliod
\cite{raboud2}). Therefore, the range of the cluster stars had to be
checked. We applied star counts in concentric rings around the centre of the
clusters.

Star counts in the vicinity of NGC\,2194 show no significant variations of the
stellar density outside a circle with a diameter of $10 \arcmin$
(corresponding to $\approx 8.5$ pc at the distance of the object) around the
centre of the cluster. For NGC\,1960, this point is more difficult to verify,
since its total stellar density is much lower than for NGC\,2194, so that it
is not as easy to see at which point a constant level is reached, and on the
other hand, its smaller distance lets us reach fainter absolute magnitudes so
that the effect of mass segregation might be more prominent within the reach
of our photometry. However, our tests provided evidence, too, that the cluster
diameter is no larger than $12 \arcmin$. It must be stressed that these
figures can only provide lower limits for the real cluster sizes: Members
fainter than the limiting magnitude of our photometry might reach further out
from the centres of the clusters.

\subsection{Proper motions}

For our proper motion studies we used photographic plates which were taken
with the Bonn Doppelrefraktor, a 30\,cm refractor ($f=5.1 \mbox{\ m}$, scale:
$40 \farcs 44 \mbox{\ mm}^{-1}$) which was located in Bonn from 1899 to 1965
and at the Hoher List Observatory of Bonn University thereafter. The 16 cm
$\times$ 16 cm plates cover a region of $1.6 ^{\circ} \times 1.6
^{\circ}$. They were completely digitized with $10 \mbox{\ } \mu \mbox{m}$
linear resolution with the Tautenburg Plate Scanner, TPS (Brunzendorf \&
Meusinger \cite{TPS}, \cite{TPS99}). The positions of the objects detected on
the photographic plates were determined using the software {\tt search} and
{\tt profil} provided by the Astronomisches Institut M\"unster (Tucholke
\cite{aim}).

In addition, we used the 1\,s to 20\,s Calar Alto exposures to improve data
quality and --- for NGC\,2194 --- to extend the maximum epoch difference.
Furthermore, a total of 16 CCD frames of NGC\,1960 which were taken with
the 1\,m Cassegrain telescope ($f/3$ with a focal reducing system) of the Hoher
List Observatory were included in the proper motion study. The latter
observations cover a circular field of view of $28 \arcmin$ in diameter which
provides a sufficiently large area for the cluster itself and the surrounding
field. The astrometric properties of this telescope/CCD camera system were
proven to be suitable for this kind of work in Sanner et
al. (\cite{holicam}). The stellar $(x,y)$ positions were extracted from the
CCD frames with DAOPHOT II routines (Stetson \cite{daophot}). A list of the
plates and Hoher List CCD images included in our study can be found in Table
\ref{caplatten}.

\begin{table}
\caption[]{\label{caplatten}Photographic plates form the Bonn Doppelrefraktor
  (prefix ``R'') and CCD frames of the 1m Cassegrain telescope of the Hoher
  List Observatory (prefix ``hl'') used to determine the proper motions of the
  stars in and around NGC\,1960 and NGC\,2194. For both clusters, the short
  ($t_{\rm exp} \leq 20 \mbox{\ s}$) Calar Alto CCD photometric data (see
  Sect. \ref{caphot} were included in the calculations, too}
\begin{tabular}{cllr}
\hline
cluster & \multicolumn{1}{c}{plate no.} & \multicolumn{1}{c}{date} &
\multicolumn{1}{c}{$t_{\rm exp} [\mbox{min}]$}\\
\hline
NGC 1960 & R0238 & 30.01.1916 & 20\\
         & R0288 & 26.01.1917 & 90\\
         & R0365 & 21.01.1919 & 240\\
         & hl01093 & 15.01.1996 & 1\\
         & hl01095 & 15.01.1996 & 1\\
         & hl01099 & 15.01.1996 & 1\\
         & hl01100 & 15.01.1996 & 1\\
         & hl01104 & 15.01.1996 & 1\\
         & hl01105 & 15.01.1996 & 1\\
         & hl01875 & 08.03.1996 & 1\\
         & hl01876 & 08.03.1996 & 1\\
         & hl01880 & 08.03.1996 & 1\\
         & hl01881 & 08.03.1996 & 1\\
         & hl01885 & 08.03.1996 & 1\\
         & hl01886 & 08.03.1996 & 1\\
         & hl02519 & 12.03.1996 & 1\\
         & hl02520 & 12.03.1996 & 1\\
         & hl02523 & 12.03.1996 & 1\\
         & hl02524 & 12.03.1996 & 1\\
         & R1902 & 15.03.1999 & 60\\
         & R1903 & 15.03.1999 & 60\\
         & R1905 & 17.03.1999 & 60\\
\hline                         
NGC 2194 & R0297 & 14.02.1917 & 20\\
         & R0333 & 14.02.1918 & 130\\
         & R0334 & 15.02.1918 & 60\\
         & R0336 & 16.02.1918 & 10\\
         & R0337 & 16.02.1918 & 60\\
         & R1125 & 01.11.1972 & 35\\
         & R1142 & 14.02.1974 & 60\\
         & R1144 & 17.02.1974 & 60\\
\hline
\end{tabular}
\end{table}

The fields of the photographic plates contain only a very limited number of
HIPPARCOS stars (ESA \cite{hipp}), as summarized in Table
\ref{hippact}. Therefore, we decided to use the ACT catalogue (Urban et
al. \cite{act}) as the basis for the transformation of the plate coordinates
$(x,y)$ to celestial coordinates $(\alpha,\delta)$. For NGC\,2194 this
decision is evident, for NGC\,1960 we preferred the ACT data, too, as the
brightest HIPPARCOS stars are overexposed on several plates, thus lowering the
accuracy of positional measurements: It turned out that only three of the
HIPPARCOS stars were measured well enough to properly derive their proper
motions from our data. The celestial positions of the stars were computed
using an astrometric software package developed by Geffert et
al. (\cite{geffert97}). We obtained good results using quadratic polynomials
in $x$ and $y$ for transforming $(x,y)$ to $(\alpha,\delta)$ for the
photographic plates and cubic polynomials for the CCD images, respectively.

Initial tests in the fields of both clusters revealed that the proper motions
computed for some ten ACT stars disagreed with the ACT catalogue values. We
assume that this is caused by the varying accuracy of the Astrographic
Catalogue which was used as the first epoch material of the ACT proper motions
or by unresolved binary stars (see Wielen et al. \cite{wielen}). We eliminated
these stars from our input catalogue.

The proper motions were computed iteratively from the individual positions:
Starting with the ACT stars to provide a calibration for the absolute proper
motions and using the resulting data as the input for the following step, we
derived a stable solution after four iterations. Stars with less than two
first and second epoch positions each or a too high error in the proper
motions ($>8 \mbox{\ mas yr}^{-1}$ in $\alpha$ or $\delta$) were not taken
into further account.

\begin{table}
\caption[]{\label{hippact} Number of HIPPARCOS and ACT stars inside the field
  of view of the Doppelrefraktor plates used for the proper motion studies}
\begin{tabular}{ccr}
\hline
cluster & HIPPARCOS & ACT \\
\hline
NGC\,1960 & 8 & 135 \\
NGC\,2194 & 4 & 82 \\
\hline
\end{tabular}
\end{table}

To determine the membership probabilities from the proper motions, we selected
$18 \arcmin$ wide areas around the centres of the clusters. This dimension
well exceeds the proposed diameter of both clusters so that we can assume to
cover all member stars for which proper motions were determined. Furthermore,
this region covers the entire field of view of the photometric data. The
membership probabilities were computed on the base of the proper motions using
the method of Sanders (\cite{sanders}): We fitted a sharp (for the members)
and a wider spread (for the field stars) Gaussian distribution to the
distribution of the stars in the vector point plot diagram and computed the
parameters of the two distributions with a maximum likelihood method. From the
values of the distribution at the location of the stars in the diagram we
derived the membership probabilities. The {\it positions} of the stars did not
play any role in the derivation of the membership probabilities. In the
following, we assumed stars to be cluster members in case their membership
probability is $0.8$ or higher.

\subsection{Colour magnitude diagrams}

\label{cacmd}

Before analysing the CMDs in detail, we had to distinguish between field and
cluster stars to eliminate CMD features which may result from the field star
population(s). For the stars down to $V=14 \mbox{\ mag}$ (NGC 1960) and $V=15
\mbox{\ mag}$ (NGC 2194) we found after cross-identifying the stars in the
photometric and astrometric measurements that our proper motion study is
virtually complete. Therefore we used these magnitudes as the limits of our
membership determination by proper motions. For the fainter stars we
statistically subtracted the field stars:

We assumed a circular region with a diameter of 806 pixels or $13 \farcm 4 $
to contain all cluster member stars. As seen in Sect. \ref{cagroesse}, this
exceeds the diameters of the clusters. The additional advantage of this
diameter of the ``cluster'' region is that this circle corresponds to exactly
half of the area covered by the CCD images so that it was not necessary to put
different weights on the star counts in the inner and outer regions. We
compared the CMDs of the circular regions containing the clusters with the
diagrams derived from the rest of the images to determine cluster CMDs without
field stars. The method is described in more detail in, e.g., Dieball \&
Grebel (\cite{dieball}).

We fitted isochrones based on the models of Bono et al. (\cite{isoteramo}) and
provided by Cassisi (private communication) to the cleaned CMDs. We assumed a
Solar metallicity of $Z=0.02$ and varied the distance modulus, reddening, and
ages of the isochrones. Comparison with the isochrones of other groups
(Schaller et al. \cite{schaller}, Bertelli et al. \cite{padua}) does not show
any significant differences in the resulting parameters.

\subsection{Mass function}

\label{caimf}

\begin{figure}
\centerline{
\includegraphics[width=\hsize]{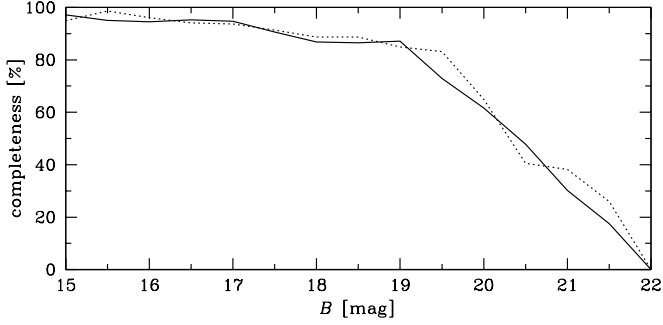}
}
\caption[]{\label{cacompl} Completeness of the $900$ s $B$ exposures of
  NGC\,1960 (solid line) and NGC\,2194 (dotted line). Down to $B=20 \mbox{\ mag}$ both samples are at least 60 \% complete. Note that the IMF is
  computed on the base of the $V$ magnitudes which alters the completeness
  function as we deal with main sequence stars with a colour of up to $B-V=1.5
  \mbox{\ mag}$}
\end{figure}

For the IMF study it is important to correct the data for the incompleteness
of our photometry. With artificial star experiments using the DAOPHOT II
routine {\tt addstar} we computed $B$-magnitude depending completeness factors
for both clusters. The $B$ photometry was favoured for these experiments since
its completeness decreases earlier as a consequence of its brighter limiting
magnitude. According to Sagar \& Richtler (\cite{sagricht}), the final
completeness of the photometry after combining the $B$ and $V$ data is well
represented by the least complete wavelength band, hence $V$ completeness was
not studied. The results, which are approximately the same for both NGC\,1960
and NGC\,2194, are plotted in Fig. \ref{cacompl}: The sample is --- except for
a few stars which likely are missing due to crowding effects --- complete down
to $B=19 \mbox{\ mag}$, and for stars with $B \la 20 \mbox{\ mag}$, we still
found more than 60 \% of the objects. In general, we found that the
completeness in the cluster regions does not differ from the values in the
outer parts of the CCD field. We therefore conclude that crowding is not a
problem for our star counts, even at the faint magnitudes. However, crowding
may lead to an increase in the photometric errors, especially in the region of
NGC\,2194, in which the stellar density is considerably higher than for
NGC\,1960.

Several objects remained far red- or bluewards of the lower part of the main sequence after statistical field star subtraction. We assume that this
results from the imperfect statistics of the sample. For a formal elimination
of these stars we had to define a region of the CMD outside of which all
objects can be considered to be non-members. This was achieved by shifting the
fitted isochrones by two times the errors listed in Table \ref{caphoterrors} in
$V$ and $B-V$ to the lower left and the upper right in the CMD (Since this
procedure applies only to stars within the range of the statistical field star
subtraction, we used the errors given for the faint stars in our
photometry.). To take into account probable double or multiple stars we added
another $0.75 \mbox{\ mag}$ to the shift to the upper right, and for NGC\,2194
we allowed another $\Delta (B-V)=0.2 \mbox{\ mag}$ in the same direction as a
consequence of the probably higher photometric errors due to crowding in the
central part of the cluster. All stars outside the corridor defined in this
way are not taken into account for our further considerations. The shifted
isochrones are plotted as dotted lines in Figs. \ref{n1960cmdhaufen} and
\ref{n2194cmdhaufen}, respectively. It may be remarked that according to Iben
(\cite{iben}) we can exclude objects with a distance of several magnitudes in
$V$ or a few tenths of magnitudes in $B-V$ from the isochrone to be pre-main
sequence members of neither NGC\,1960 nor NGC\,2194.

We furthermore selected all objects below the turn-off point of the
isochrones. For the remaining stars, we calculated their initial masses on
the base of their $V$ magnitudes. We used the mass-luminosity relation
provided with the isochrone data. $V$ was preferred for this purpose as the
photometric errors are smaller in $V$ compared to the corresponding $B$
magnitudes. The mass-luminosity relation was approximated using
$6^{\mbox{th}}$ order polynomials
\begin{equation} \label{camlreq}
m[M_\odot]=\sum_{i=0}^6 a_i \cdot (V[\mbox{mag}])^i
\end{equation}
which resulted in an
rms error of less than 0.01. Using $5^{\mbox{th}}$ or lower order polynomials
caused higher deviations especially in the low mass range. The values of the
parameters $a_i$ are listed in Table \ref{camlr}.

\begin{table}
\caption[]{\label{camlr}Parameters of the polynomial approximation of the
  mass-luminosity relation for the stars of the two clusters. See
  Eq. (\ref{camlreq}) for the definition of $(a_0, \ldots, a_6)$}
\begin{tabular}{rr@{$\cdot$}lr@{$\cdot$}l}
\hline
 & \multicolumn{2}{c}{NGC\,1960} & \multicolumn{2}{c}{NGC\,2194} \\
\hline
$a_0$ & $-3.0892$ & $10^{+2}$ & $+1.2671$ & $10^{+2}$ \\
$a_1$ & $+1.4661$ & $10^{+2}$ & $-3.5052$ & $10^{+1}$ \\
$a_2$ & $-2.6277$ & $10^{+1}$ & $+4.3076$ & $10^{0}$ \\
$a_3$ & $+2.3780$ & $10^{0}$ & $-2.9135$ & $10^{-1}$ \\
$a_4$ & $-1.1671$ & $10^{-1}$ & $+1.1218$ & $10^{-2}$ \\
$a_5$ & $+2.9728$ & $10^{-3}$ & $-2.3033$ & $10^{-4}$ \\
$a_6$ & $-3.0881$ & $10^{-5}$ & $+1.9556$ & $10^{-6}$ \\
\hline
\end{tabular}
\end{table}

\begin{figure}
\centerline{
\includegraphics[width=\hsize]{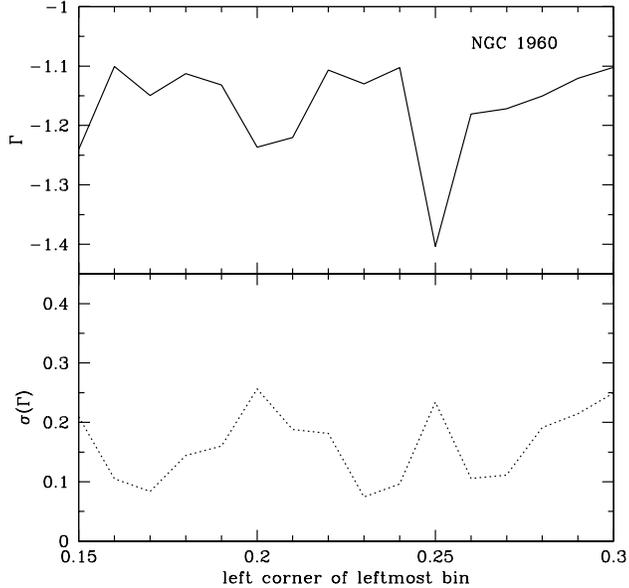}
}
\caption[]{\label{n1960binshift} Results of an experiment with
part of the NGC\,1960 data. We computed histograms with an equal bin width of
$\Delta \log m[M_\odot]=0.1$ (as in Fig. \ref{n1960imfplot}), but varying
location of the bins, expressed by the $\log m$ value of the left corner of
the leftmost bin. This diagram shows the resulting IMF slope $\Gamma$ and
the corresponding error of the fit}
\end{figure}

Taking into account the incompleteness of the data, we determined the
luminosity and initial mass functions of the two clusters. The IMF slope was
computed with a maximum likelihood technique. We preferred this method instead
of the ``traditional'' way of a least square fit of the mass function to a
histogram, because those histogram fits are not invariant to size and location
of the bins: Experiments with shifting the location and size of the bins
resulted in differences of the exponent of more than $\Delta
\Gamma=0.2$. Fig. \ref{n1960binshift} shows the results of such an experiment
with the NGC\,1960 data. The fitted IMFs show an average $\Gamma$ value of
around $-1.2$ with individual slopes ranging from $-1.1$ down to $-1.4$. This
can be explained by the very small number of stars in the higher mass bins
which contain only between one and ten stars. In case only one member is
mis-interpreted as a non-member or vice versa, the corresponding bin height
might be affected by up to $\Delta \log N(m)=0.3$ in the worst case which will
heavily alter the corresponding IMF slope. In addition, all {\it bins} of the
histogram obtain the same weight in the standard least square fit, no matter
how many stars are included. For very populous or older objects (globular or
older open star clusters, see, e.g., the IMF of NGC\,2194) this effect plays a
minor role, because in these cases the number of stars per bin is much
higher. On the other hand, the maximum likelihood method does not lose
information (as is done by binning), and each {\it star} obtains the same
weight in the IMF computation.

Nevertheless, for reasons of illustration we sketch the IMF of our two
clusters with an underlying histogram in Figs. \ref{n1960imfplot} and
\ref{n2194imfplot}.

\section{NGC\,1960}

\label{n1960disc}

\subsection{Proper motion study}

\label{n1960eb}

With the method described above, we determined the proper motions of 1,190
stars within the entire field of the photographic plates. We found that the
limiting magnitude of the second epoch plates is brighter than that of the
first epoch plates. This effect is compensated by the addition of the CCD
data, so that in the cluster region we reach fainter stars than in the outer
region of the field. Therefore, the limiting magnitude of the proper motion
study is fainter in the area for which the CCD data were available.

After four iterations of proper motion determination, the comparison of the
computed proper motions with ACT led to systematic positional differences of
the order of $\Delta \alpha=0 \farcs 04$ and $\Delta \delta=0 \farcs 10$ and
for the proper motions of around $0.4 \mbox{\ mas}$ $\mbox{yr}^{-1}$. The internal
dispersion of the proper motions was computed  from the deviations of the
positions from a linear fit of $\alpha$ and $\delta$ as functions of time. We
derived mean values of $\sigma_{\mu_\alpha \cos   \delta}=1.1 \mbox{\ mas
  yr}^{-1}$ and $\sigma_{\mu_\delta}=0.9 \mbox{\ mas yr}^{-1}$ for individual
stars.

\begin{figure}
\centerline{
\includegraphics[width=\hsize]{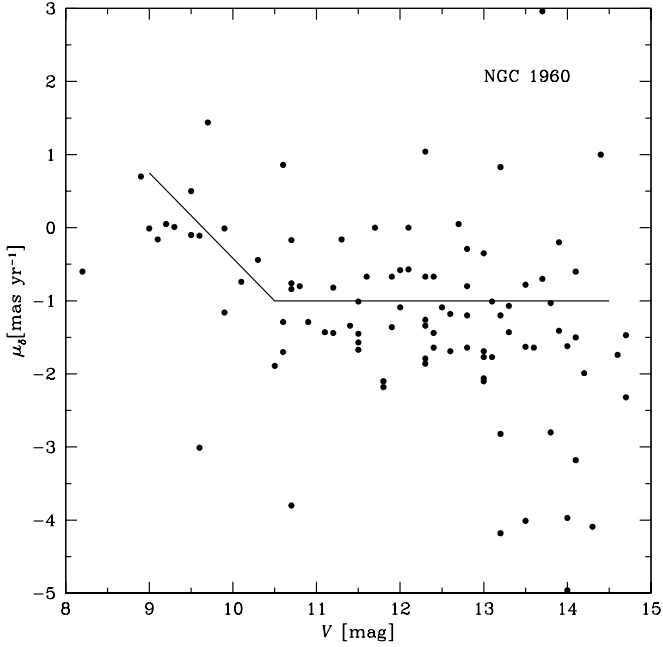}
}
\caption[]{\label{n1960gsc}Dependence of the proper motions in $\delta$ on the
  magnitude for the stars in the field of NGC\,1960. The diagram shows a clear
  magnitude term for the stars brighter than 10.5 mag. Note that these proper
  motions were computed on the basis of GSC 1.2 so that we deal with {\it
  relative} proper motions here. The solid line sketches the mean proper
  motion for the cluster member stars. See the text for further discussion}
\end{figure}

We detected a slight, but systematic slope of the proper motions in
$\delta$ depending on the magnitude of the stars resulting in a difference of
approximately $2 \mbox{\ mas yr}^{-1}$ for the proper motions between the
brightest and faintest members. As the magnitude range of the ACT catalogue
within our field of view is limited to approximately 11 mag, we used the
positions provided by the Guide Star Catalog (GSC) Version 1.2 (R\"oser et
al. \cite{gsc}), which covers stars over the entire range of our proper motion
study, for further analysis. The disadvantage of GSC is the fact that it does
not provide proper motions. Therefore, all results obtained with this
catalogue are {\it relative} proper motions only. Fig. \ref{n1960gsc} shows
the proper motions in $\delta$ and their dependence on the magnitudes derived
from this computation for the stars in the inner region of the photographic
plates. The diagram shows that only the stars brighter than $10.5 \mbox{\ mag}$
are influenced by a clear magnitude term leading to deviations of up to $2
\mbox{\ mas yr}^{-1}$ with respect to the stars fainter than $10.5 \mbox{\
  mag}$ which do not show any systematic trend. We included a magnitude term
into our transformation model, however, since the behaviour is not linear with
magnitudes and different from plate to plate we were unable to completely
eliminate the effect. Furthermore, taking into account that many of the ACT
stars are brighter than $10.5 \mbox{\ mag}$, it is clear that this deviation
was extrapolated over the entire range of the proper motion study.

Meurers (\cite{meurers}), who had used the same first epoch material for his
proper motion study, found a similar phenomenon and suggested that the bright
and the faint stars in the region of NGC\,1960 form independent stellar
``aggregates''. In his study the proper motion difference between bright and
faint stars is much more prominent. Taking into account his smaller epoch
difference of 36 years this could be explained assuming that the effect is
caused by (at least some of) the first epoch plates on which the positions of
the brighter stars seem to be displaced by an amount of approximately $0.1
\arcsec$ to $0.2 \arcsec$ compared to the fainter objects, whereas both his
and our second epoch data are unaffected. This proposition would also explain
why we did not detect this inaccuracy during the determination of the
positions on the plates, since the uncertainties of single positional
measurements are of the same order of magnitude.

The proper motions in $\alpha$ proved to be unaffected by this phenomenon.

We found that when using the ACT based proper motions, the membership
determination  is not affected by this problem, since the magnitude trend in
$\mu_\delta$ is smoothed over the magnitude range (in comparison with
Fig. \ref{n1960gsc}). On the other hand, in the GSC solution, the bright stars
have proper motions differing too much from the average so that almost all of
them are declared non-members. Therefore we used the results based on the ACT
data for the computation of the membership probabilities. Table
\ref{n1960cdseb} shows a list of all proper motions determined on the base of
the ACT catalogue.

\begin{table}
\caption[]{\label{n1960cdseb} List of all stellar proper motions determined
  from the photographic plates and the additional CCD images of NGC\,1960. The
  positions are given for the epoch 1950.0 in the equinox J2000.0 coordinate
  stystem. The stellar id numbers are the same as in Table
  \ref{n1960cdsphot}. Again, the stellar numbers from Boden (\cite{boden}) are
  listed in addition. Only the proper motions of the same stars as in Table
  \ref{n1960cdsphot} are presented here, the complete table is available
  online at the CDS archive in Strasbourg}
\begin{tabular}{rrrrrr}
\hline
\multicolumn{1}{c}{No.} & \multicolumn{1}{c}{Boden} & \multicolumn{1}{c}{$\alpha_{2000}$} & \multicolumn{1}{c}{$\delta_{2000}$} & \multicolumn{1}{c}{$\mu_\alpha \cos \delta$} & \multicolumn{1}{c}{$\mu_\delta$} \\
 & \multicolumn{1}{c}{No.} & \multicolumn{1}{c}{[hhmmss.sss]} & \multicolumn{1}{c}{[$^\circ$ $\arcmin$
   $\arcsec$]} & \multicolumn{2}{c}{[mas yr$^{-1}$]}\\
\hline
$ 1$ & $   $ & $053615.805$ & $+340836.81$ & $+2.29$ & $-7.78$\\
$ 2$ & $ 23$ & $053623.059$ & $+341032.84$ & $+2.28$ & $-8.46$\\
$ 3$ & $138$ & $053639.262$ & $+340349.88$ & $+2.33$ & $-9.55$\\
$ 4$ & $101$ & $053642.305$ & $+341206.09$ & $+3.83$ & $-8.43$\\
$ 5$ & $ 61$ & $053632.004$ & $+341047.44$ & $+2.36$ & $-8.59$\\
$ 8$ & $ 27$ & $053628.031$ & $+340830.78$ & $+3.36$ & $-8.02$\\
$ 9$ & $ 48$ & $053607.902$ & $+340850.53$ & $+1.16$ & $-8.36$\\
$10$ & $ 21$ & $053621.062$ & $+341016.91$ & $+2.46$ & $-8.35$\\
$11$ & $ 38$ & $053623.844$ & $+340557.38$ & $+2.45$ & $-8.61$\\
$12$ & $184$ & $053546.469$ & $+340317.47$ & $-5.30$ & $-6.61$\\
\multicolumn{1}{c}{$\vdots$} & \multicolumn{1}{c}{$\vdots$} &
\multicolumn{1}{c}{$\vdots$} & \multicolumn{1}{c}{$\vdots$} &
\multicolumn{1}{c}{$\vdots$} & \multicolumn{1}{c}{$\vdots$}\\
\hline
\end{tabular}
\end{table}

\begin{figure}
\centerline{
\includegraphics[width=\hsize]{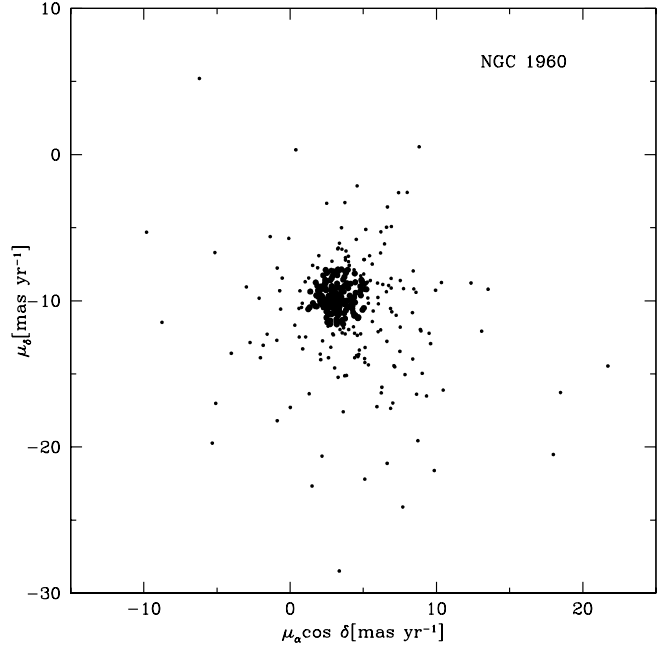}
}
\caption[]{\label{n1960vppd} Vector point plot diagram of the stars in the
  region of NGC\,1960. The stars with a membership probability of less than
  0.8 are indicated by small, the others by larger dots. The width of the
  distribution of the stars with a high membership probability is of the order
  of $1 \mbox{\ mas\,yr}^{-1}$ which coincides with the standard deviation of
  the proper motion of a single star}
\end{figure}

\begin{figure}
\centerline{
  \includegraphics[width=\hsize]{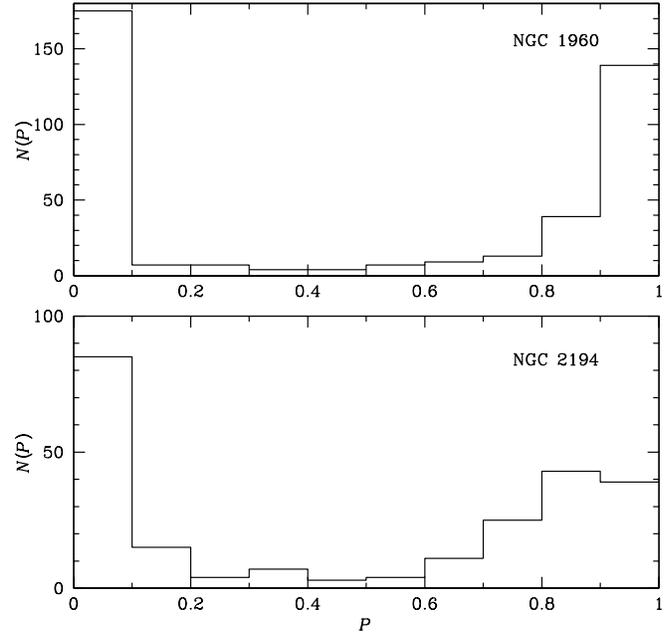}
}
\caption[]{\label{caphist} Histograms of the membership probabilities for the
  stars of NGC\,1960 (upper diagram) and NGC\,2194 (lower diagram). All stars
  with membership probabilities of 0.8 or higher are considered cluster
  members. Note that the separation between members and non-members is less
  prominent for NGC\,2194 than for NGC\,1960}
\end{figure}

The vector point plot diagram as determined on the base of ACT for the stars
in the central region of the plates is presented in
Fig. \ref{n1960vppd}. Membership determination resulted in 178 members and 226
non-members of NGC\,1960. The distribution of membership probabilities
sketched in Fig. \ref{caphist} shows a clear separation of members and
non-members with only a small number of stars with intermediate membership
probabilities. The centre of the proper motion distribution of the cluster
members in the vector point plot diagram is determined to be
\begin{eqnarray}
\mu_\alpha \cos \delta &=& 3.2 \pm 1.1 \mbox{\ mas yr}^{-1} \mbox{\ and } \label{actebalpha}\\
\mu_\delta &=& -9.8 \pm 0.9 \mbox{\ mas yr}^{-1}. \label{actebdelta}
\end{eqnarray}
The width of the Gaussian distribution of the proper motions is around $1 \mbox{\ mas yr}^{-1}$ and hence the same as the $1 \sigma$ error of the proper
motion of a single object. The field moves very similarly with 
\begin{eqnarray}
\mu_\alpha \cos \delta &=& 4.4 \pm 5 \mbox{\ mas yr}^{-1} \mbox{\ and }\\
\mu_\delta &=& -11.4 \pm 5 \mbox{\ mas yr}^{-1}.
\end{eqnarray}
The similarity of field and cluster proper motions makes membership
determination a difficult task: Several field stars which by chance have the
same proper motion as the cluster stars will be taken for members.

These results cannot be used for a determination of the {\it absolute} proper
motion of the cluster, since the centre of the distribution can be assumed to
be displaced upwards in the vector point plot diagram as a consequence of the
magnitude dependence of the $\mu_\delta$ values. To obtain reliable absolute
proper motions, nevertheless, we used the fainter ($> 10.5 \mbox{\ mag}$) part
of the proper motions computed on the base of GSC 1.2 which are stable with
magnitudes and compared their relative proper motions with the values given
for the corresponding stars in the ACT. We found a difference of $\Delta
(\mu_\alpha \cos \delta)=1.4 \pm 2.6$ and $\Delta \mu_\delta=-6.5 \pm 2.4$ and
centres of the GSC based proper motion distributions of $\mu_\alpha \cos
\delta=1.5 \pm 0.7$ and $\mu_\delta=-1.5 \pm 0.7$. As a consequence we
determined the absolute proper motions of NGC\,1960 to be
\begin{eqnarray}
\mu_\alpha \cos \delta &=& 2.9 \pm 2.7 \mbox {\ mas yr}^{-1} \mbox{\ and }\\
\mu_\delta &=& -8.0 \pm 2.5 \mbox {\ mas yr}^{-1}.
\end{eqnarray}
As expected, the value of the proper motion in right ascension is --- compared
to Eq. (\ref{actebalpha}) --- unaffected within the errors, whereas
$\mu_\delta$ is different from the value of Eq. (\ref{actebdelta}) by a value
which corresponds to the $2 \sigma$ error of $\mu_\delta$.

\subsection{Colour magnitude diagram properties}

\label{n1960cmdsect}

\begin{figure}
\centerline{
\includegraphics[width=\hsize]{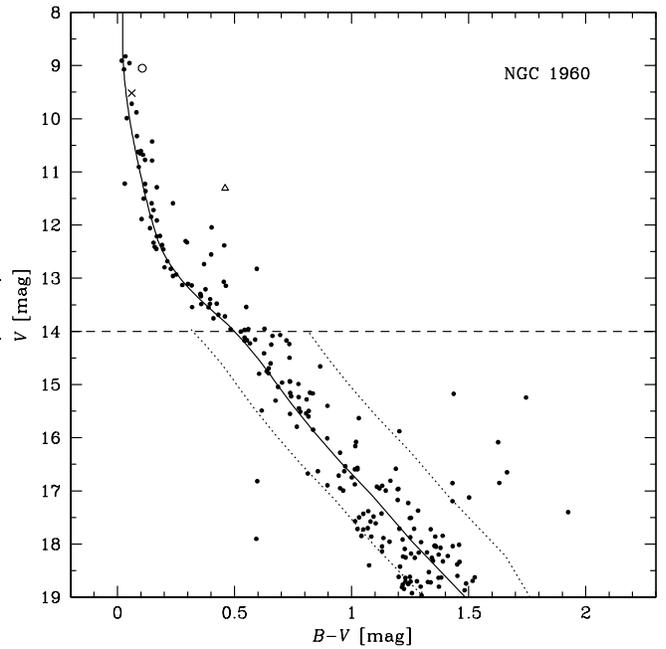}
}
\caption[]{\label{n1960cmdhaufen} Colour magnitude diagram of all members of
  NGC\,1960 as determined with the proper motions ($V<14 \mbox{\ mag}$) and
  statistical field star subtraction ($V>14 \mbox{\ mag}$). The dashed line
  stands for the borderline between the two methods of membership
  determination; the dotted lines indicate the corridor containing the main
  sequence for the IMF computation (see Sect. \ref{caimf} for details). The
  parameters of the isochrone plotted in the diagram are listed in Table
  \ref{caparams}. The three stars marked with special symbols are discussed in
  Sect. \ref{n1960cmdsect}}
\end{figure}

The CMD of NGC\,1960 (Fig. \ref{n1960cmdhaufen}) shows a clear and narrow main
sequence with an indication of a second main sequence including approximately
15 stars from $V=11.5 \mbox{\ mag}$ to $V=14 \mbox{\ mag}$ (corresponding to
masses from $m \approx 3.5 M_\odot$ to $m \approx 1.4 M_\odot$). These stars
might be Be stars (see, e.g., Zorec \& Briot \cite{zorec}) or unresolved
binaries (see, e.g., Abt \cite{abt} or the discussion in Sagar \& Richtler
\cite{sagricht}).

Slettebak (\cite{slettebak}) reports on two Be stars in NGC 1960. One of them,
our star 1374 (Boden's (\cite{boden}) star No. 505, erroneously named No. 504
by Slettebak), clearly fails to fulfil our membership criterion with a proper
motion of $\mu_\alpha \cos \delta=-0.7 \mbox{\ mas yr}^{-1}$ and
$\mu_\delta=-0.3 \mbox{\ mas yr}^{-1}$. In addition, it is located so far off
the centre of the cluster that it is even outside the field of our CCD
images. On the other hand, star 4 (Boden's (\cite{boden}) star No. 101,
$V=9.050 \mbox{\ mag}$, $B-V=0.106 \mbox{\ mag}$) shows a proper motion of
$\mu_\alpha \cos \delta=3.9 \mbox{\ mas yr}^{-1}$ and $\mu_\delta=-8.8 \mbox{\
  mas yr}^{-1}$ which makes it a very likely cluster member with a membership
probability of $0.95$. In Fig. \ref{n1960cmdhaufen}, this object is marked
with a circle. A third Be star in the region is mentioned in the WEBDA
database (Mermilliod \cite{webda}): Boden's (\cite{boden}) star No. 27, or our
star 8. We obtained a proper motion of $\mu_\alpha \cos \delta=3.36 \mbox{\
  mas yr}^{-1}$, $\mu_\delta=-8.02 \mbox{\ mas yr}^{-1}$. From these figures
we computed a membership probability of $0.89$ so that this star is a likely
cluster member. We marked this object in Fig. \ref{n1960cmdhaufen} with a
cross. As there is no evidence for any further Be stars, it is plausible to
assume that the other stars forming the second main sequence most likely are
unresolved binary stars.

The star at $V=11.3 \mbox{\ mag}$, $B-V=0.46 \mbox{\ mag}$ (star 29 of our
sample, marked with a triangle in Fig. \ref{n1960cmdhaufen}) does not fit to
any isochrone which properly represents the other stars in this magnitude
range. It shows a proper motion of $\mu_\alpha \cos \delta=2.9 \mbox{\ mas
  yr}^{-1}$ and $\mu_\delta=-9.7 \mbox{\ mas yr}^{-1}$ resulting in a
membership probability of $0.97$. This object may be an example for a star
which coincidentally shows the same proper motion as the cluster, but being in
fact a non-member.

From our isochrone fit, we derived the parameters given in
Table \ref{caparams}. Age determination was quite a difficult task for
NGC\,1960, as there are no significantly evolved (red) stars present in
the CMD. We found that the 16 Myr isochrone might be the optimal one, since
it represents the brightest stars better than the (in terms of ages)
neighbouring isochrones. The comparably large error we adopted reflects this
consideration.

\begin{table}
\caption{\label{caparams} Parameters of NGC\,1960 and NGC\,2194 as derived
  from isochrone fitting to the colour magnitude diagram}
\begin{tabular}{lr@{ = }l}
\hline
\multicolumn{3}{c}{NGC\,1960}\\
\hline
distance modulus & $(m-M)_0$ & $ 10.6 \pm 0.2 \mbox{\ mag}$\\
i.e. distance    & $r$       & $ 1318 \pm 120 \mbox{\ pc}$\\
reddening        & $E_{B-V}$ & $ 0.25 \pm 0.02 \mbox{\ mag}$\\
age              & $t$  & $ 16^{+10}_{-5} \mbox{\ Myr}$\\
i.e.             & $\log t$  & $ 7.2 \pm 0.2$\\
metallicity      & $Z$       & $0.02$\\
\hline
\multicolumn{3}{c}{NGC\,2194}\\
\hline
distance modulus & $(m-M)_0$ & $12.3 \pm 0.2 \mbox{\ mag}$\\
i.e. distance    & $r$       & $ 2884 \pm 270 \mbox{\ pc}$\\
reddening        & $E_{B-V}$ & $0.45 \pm 0.02 \mbox{\ mag}$\\
age              & $t$       & $ 550 \pm 50 \mbox{\ Myr}$\\
i.e.             & $\log t$  & $ 8.74 \pm 0.05$\\
metallicity      & $Z$       & $0.02$\\
\hline
\end{tabular}
\end{table}

\subsection{Initial mass function}

\begin{figure}
\centerline{
\includegraphics[width=\hsize]{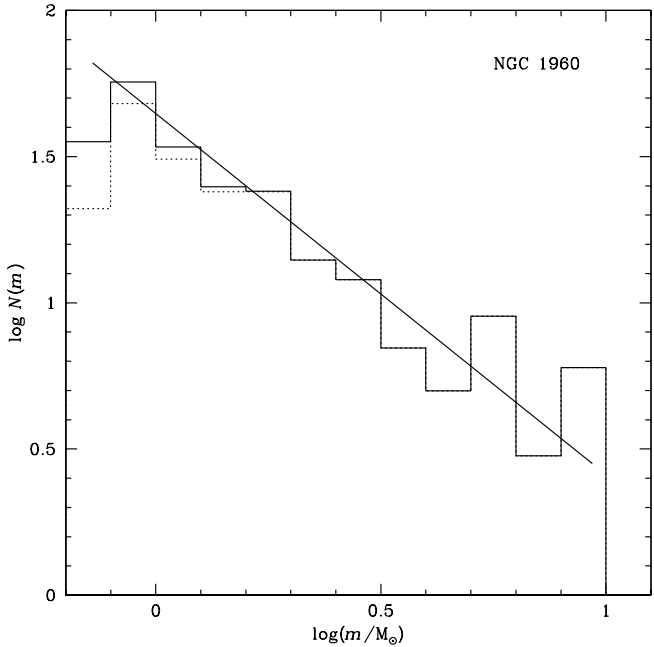}
}
\caption[]{\label{n1960imfplot} Initial mass function of NGC\,1960.
  The solid histogram corresponds to the completeness corrected values, the
  dotted one to the original data. The IMF slope calculated with a maximum
  likelihood analysis is determined to be $\Gamma=-1.23 \pm 0.17$. The stars
  with $m < 0.725 M_\odot$ (i.e. $\log (m/M_\odot)<-0.14$) were not taken into
  account for the slope determination as the average completeness for those
  stars is below 60 \%. The limits of the IMF line illustrate the mass range
  under consideration. The probable binary nature of some objects is not
  taken into account in this plot}
\end{figure}

The determination of the IMF slope from the completeness corrected data
obtained from the CMD (Fig. \ref{n1960cmdhaufen}) leads to the value of
$\Gamma=-1.23 \pm 0.17$ for NGC\,1960 in a mass interval from $m=9.4 M_\odot$
down to $m=0.725 M_\odot$ (corresponding to $V=8.9 \mbox{\ mag}$ to $V=19
\mbox{\ mag}$). This restriction was chosen to guarantee a completeness of the
photometry of at least 60 \%. To test the stability of the IMF concerning the
probable double star nature of several objects, we assumed the stars above the
brighter part of the main sequence (a total of 18 objects) to be unresolved
binary stars with a mass ratio of 1 and computed the IMF of this modified
sample, as well. The slope increased to the value of $\Gamma=-1.19 \pm 0.17$
within the same mass range, representing a slightly shallower IMF. Anyway, the
influence of a binary main sequence is negligible within the errors. We also
experimented with leaving out the magnitude range critical for membership
determination (see Sect. \ref{n1960eb} and Fig. \ref{n1960gsc}), i.e. the
stars brighter than $V=10.5 \mbox{\ mag}$ ($m>5.75 M_\odot$), and derived
$\Gamma=-1.26 \pm 0.2$ --- a result which coincides well within the errors
with the above ones. This shows once more that the membership determination
--- and therefore the IMF --- was almost not affected by the magnitude term of
our proper motion study. Fig. \ref{n1960imfplot} sketches the IMF of
NGC\,1960.

\begin{figure}
\centerline{
\includegraphics[width=\hsize]{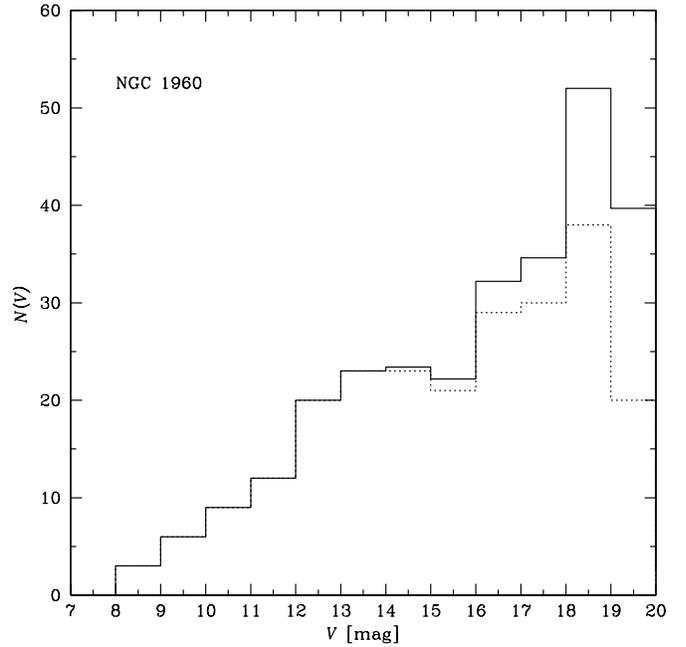}
}
\caption[]{\label{n1960lkf} Luminosity function of NGC\,1960. The solid line
  stands for the completeness corrected data, the dotted line for the
  uncorrected values. The rightmost bin almost reaches the limit of our
  photometric study so that it cannot be taken for reliable}
\end{figure}

\section{NGC\,2194}

\label{n2194disc}

\subsection{Proper motion study}

\label{n2194pm}

For stars brighter than $V=9 \mbox{\ mag}$, some of the plates of NGC 2194
showed a systematic shift of the computed $\delta$ positions with respect to
the ACT values. We therefore excluded all those stars from our input
catalogue. However, this effect --- which is different from the one described
before for NGC\,1960, since it perceptibly affects the positions --- does not
influence our membership determination as the region of NGC\,2194 does not
cover any stars of this brightness (see also Table \ref{n2194cdsphot}).

\begin{figure}
\centerline{
\includegraphics[width=\hsize]{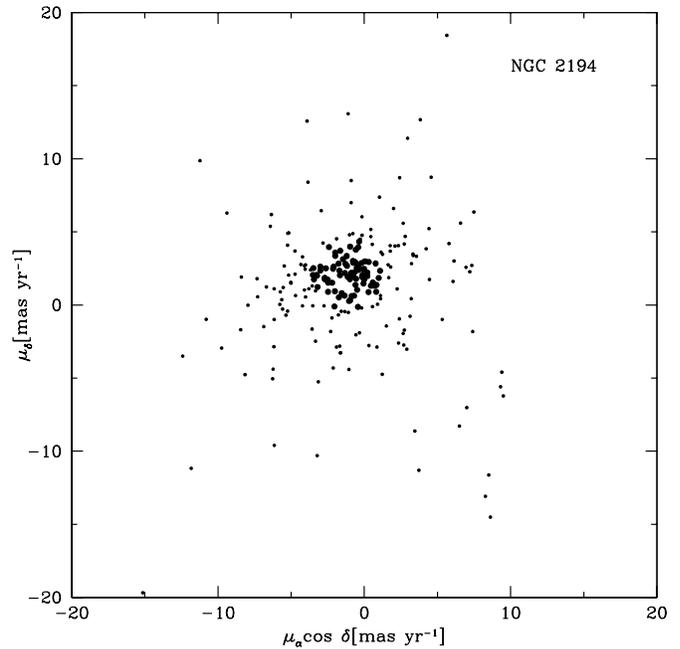}
}
\caption[]{\label{n2194vppd} Vector point plot diagram of the stars in the
  region of NGC\,2194. As in Fig. \ref{n1960vppd}, the stars with a membership
  probability of higher than 0.8 are indicated by large, the others by small
  dots. The single errors are of the order of $1.5 \mbox{\ mas yr}^{-1}$}
\end{figure}

Proper motions of 2,233 stars could be computed from the plates of
NGC\,2194. This figure is significantly higher than for NGC\,1960, since this
time the second epoch plates are of much higher quality, so that we reach
fainter stars over the entire $1.6^{\circ} \times 1.6^{\circ}$ field. 
After four iterations of our proper motion determination, the systematic
difference between ACT and our results were $0 \farcs 07$ for the positions
and $0.07 \mbox{\ mas yr}^{-1}$ for the proper motions. The standard deviation
of the proper motions were $\sigma_{\mu_\alpha \cos \delta}=1.6 \mbox{\ mas
  yr}^{-1}$ and $\sigma_{\mu_\delta}=1.5 \mbox{\ mas yr}^{-1}$. The vector
point plot diagram (Fig. \ref{n2194vppd}) of the stars in the region of
NGC\,2194 shows that the stars are not as highly concentrated in the centre of
the distribution of the cluster stars as for NGC\,1960. This also explains the
less distinct peak for high membership probabilities in the histogram shown in
Fig. \ref{caphist}.

Although one finds more stars on the whole plates, in the inner region the
proper motions of fewer objects were detected. This is caused by the lower
number of sufficiently bright stars in and around the cluster (see
Figs. \ref{n2194bild} and \ref{n2194cmd}). On the other hand, the low part of
the main sequence is much more densely populated. We will see in
Sect. \ref{n2194cmdsect} that the total number of members detected is higher
for NGC\,2194 than for NGC\,1960.

We classified 149 members and 81 non-members. For this cluster, the
separation between members and non-members was even more difficult as can be
seen from the membership probability histogram plotted in
Fig. \ref{caphist}: Approximately 50 stars show intermediate membership
probabilities between 0.2 and 0.8. The result for the absolute proper motion
of the cluster is
\begin{eqnarray}
\mu_\alpha \cos \delta &=& -2.3 \pm 1.6 \mbox{\ mas yr}^{-1} \mbox{\ and }\\
\mu_\delta &=& 0.2 \pm 1.5 \mbox{\ mas yr}^{-1}
\end{eqnarray}
and for the field
\begin{eqnarray}
\mu_\alpha \cos \delta &=& -0.4 \pm 5.5 \mbox{\ mas yr}^{-1} \mbox{\ and }\\
\mu_\delta &=& 0.1 \pm 5.5 \mbox{\ mas yr}^{-1}.
\end{eqnarray}
The measured standard deviation of the cluster proper motion
distribution again is the same as was determined for one individual object. 

Table \ref{n2194cdseb} shows a list of all proper motions computed.

\begin{table}
\caption[]{\label{n2194cdseb} List of all stellar proper motions determined
  from the photographic plates of NGC\,2194. Besides our internal numbering
  system, the numbers of del\,Rio's (\cite{delrio}) study are given. For more
  information see Table \ref{n1960cdseb}. The complete table is available
  online at the CDS archive in Strasbourg}
\begin{tabular}{rcrrrr}
\hline
\multicolumn{1}{c}{No.} & del\,Rio & \multicolumn{1}{c}{$\alpha_{2000}$} & \multicolumn{1}{c}{$\delta_{2000}$} & \multicolumn{1}{c}{$\mu_\alpha \cos \delta$} & \multicolumn{1}{c}{$\mu_\delta$} \\
 & \multicolumn{1}{c}{No.} & \multicolumn{1}{c}{[hhmmss.sss]} & \multicolumn{1}{c}{[$^\circ
   \mbox{\ } \arcmin \mbox{\ } \arcsec$ ]} & \multicolumn{2}{c}{[mas yr$^{-1}$]}\\
\hline
$ 1$ & $   $  & $061346.145$ & $+124352.43$ &$-88.08$ & $+19.39$\\
$ 2$ & $   $  & $061315.582$ & $+124525.58$ & $+2.46$ & $ -1.61$\\
$ 3$ & $   $  & $061357.445$ & $+125258.92$ & $+9.64$ & $-14.54$\\
$ 4$ & $ 39$  & $061350.910$ & $+124736.25$ & $-0.43$ & $ +1.65$\\
$ 5$ & $   $  & $061318.016$ & $+124113.11$ & $-4.77$ & $ -1.43$\\
$ 6$ & $   $  & $061403.945$ & $+124446.31$ & $-1.88$ & $ -2.28$\\
$ 7$ & $ 49$  & $061348.746$ & $+124823.41$ & $-4.13$ & $ +5.11$\\
$ 8$ & $   $  & $061342.734$ & $+124158.29$ & $-0.16$ & $ -2.58$\\
$10$ & $ 14$  & $061341.836$ & $+124806.10$ & $-4.12$ & $ +4.38$\\
$11$ & $   $  & $061325.805$ & $+124431.25$ & $+0.92$ & $ -0.08$\\
\multicolumn{1}{c}{$\vdots$} & \multicolumn{1}{c}{$\vdots$} &
\multicolumn{1}{c}{$\vdots$} & \multicolumn{1}{c}{$\vdots$} &
\multicolumn{1}{c}{$\vdots$} & \multicolumn{1}{c}{$\vdots$}\\
\hline
\end{tabular}
\end{table}

\subsection{Colour magnitude diagram properties}

\label{n2194cmdsect}

\begin{figure}
\centerline{
\includegraphics[width=\hsize]{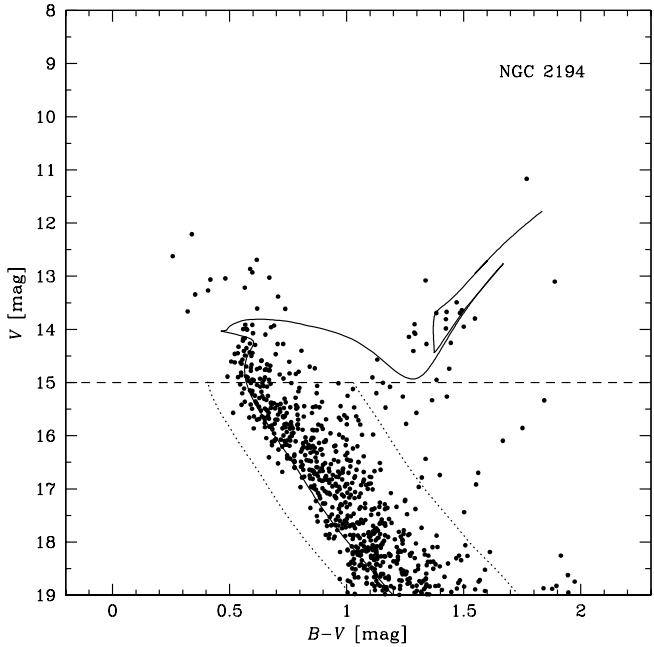}
}
\caption[]{\label{n2194cmdhaufen} Colour magnitude diagram of all members of
  NGC\,2194 as determined with the proper motions ($V<15 \mbox{\ mag}$) and
  the statistical field star subtraction ($V>15 \mbox{\ mag}$). For more
  information see Fig. \ref{n1960cmdhaufen}}
\end{figure}

According to the field star subtracted CMD, presented as
Fig. \ref{n2194cmdhaufen}, NGC\,2194 shows a prominent main sequence with a
turn-off point near $V=14.5 \mbox{\ mag}$ and a sparsely populated red giant
branch. For this cluster, the proper motion study is not of great value for
the isochrone fitting process: As the main sequence turn-off is located around
$V=14.5 \mbox{\ mag}$, it is clear that the bright blue stars either do not
belong to the cluster or do not help in finding the best isochrone fit because
of their non-standard evolution like blue stragglers (see, e.g., Stryker
\cite{stryker}). We assume that the presence of the blue bright stars is
mainly caused by the coincidence of the field and cluster proper motion
centres which causes a certain number of fields stars to be mis-identified as
cluster members. As expected in Sect. \ref{n2194pm}, this effect is more
dramatic here than in the case of NGC\,1960. From the comarison of the
isochrones we derived the parameters for NGC\,2194 given in Table
\ref{caparams}.

\begin{figure}
\centerline{
\includegraphics[width=\hsize]{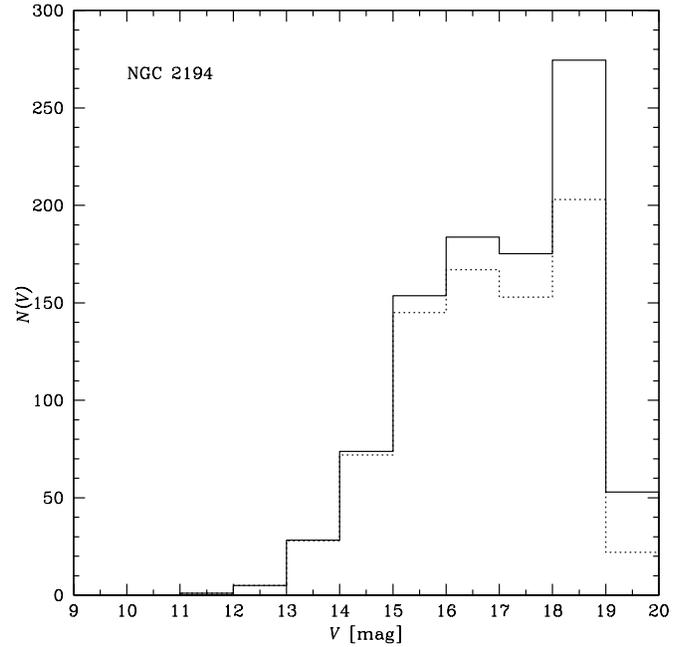}
}
\caption[]{\label{n2194lkf} Luminosity function of NGC\,2194. As before, the
  solid line stands for the completeness corrected data, the dotted line for
  the uncorrected values. Again, the rightmost bin touches the limiting
  magnitude of our photometry}
\end{figure}

Star No. 38 of our sample was first mentioned by del\,Rio (\cite{delrio},
star 160 therein) who considers this object a field star as a consequence of
its location in the CMD. For the same reason, but with the opposite
conclusion, Ahumada \& Lapasset (\cite{bluestrag}) mention this object as a
cluster member in their ``Catalogue of blue stragglers in open star
clusters''. We find for this star a proper motion of $\mu_\alpha \cos \delta=2.5 \mbox{\ mas yr}^{-1}$ and $\mu_\delta=-0.1 \mbox{\ mas yr}^{-1}$
leading to a membership probability of 0.34. Therefore, we agree with del\,Rio
and assume star 38 to be a field star, too. So far, no further information is
known about the bright blue stars in Fig. \ref{n2194cmdhaufen}, so that we
cannot give any definite statement about the nature of these objects.

\subsection{Initial mass function}

The age of NGC\,2194 of $550$ Myr --- together with its distance of almost $3$
kpc --- implies that the range of the observable main sequence is very
limited. We therefore could compute the IMF only over the mass interval from
$m \approx 1 M_\odot$ (corresponding to $V \approx 19 \mbox{\ mag}$) to $m
\approx 2.1 M_\odot$ (or $V \approx 15.0 \mbox{\ mag}$). The slope determined
is $\Gamma=-1.33 \pm 0.29$. The comparably higher error is a consequence of
the smaller mass interval.

\begin{figure}
\centerline{
\includegraphics[width=\hsize]{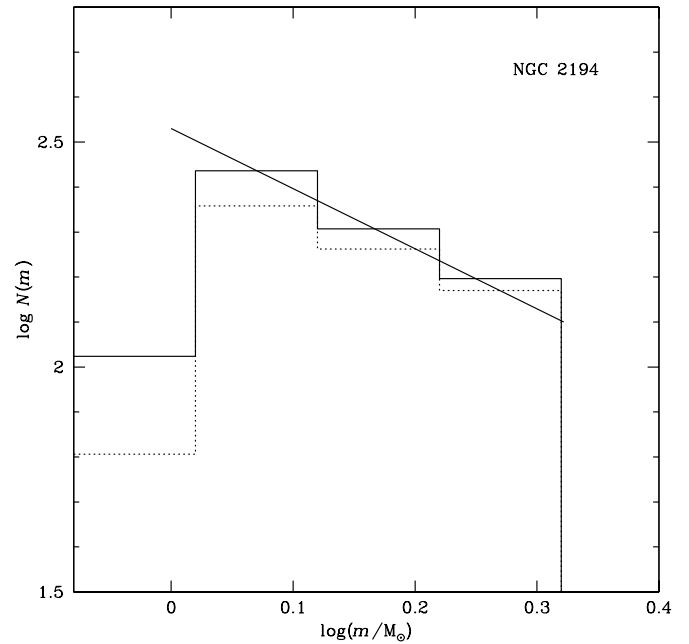}
}
\caption[]{\label{n2194imfplot} Initial mass function of NGC\,2194. The solid
  histogram corresponds to the completeness corrected values, the dotted one
  to the original data. The IMF slope is determined to be $\Gamma= -1.33 \pm
  0.29$. Note that the slope was determined with a maximum likelihood
  analysis; the histogram is only plotted for demonstration. The length of the
  line illustrates the mass interval for which the IMF was computed}
\end{figure}

Comparing the resulting IMF with a histogram (see Fig. \ref{n2194imfplot}),
one finds good agreement for three of the four bins (The bin width
is the same as for NGC\,1960). As the completeness drops rapidly within the
leftmost bin, it has --- in total --- a high uncertainty. However, only stars
with masses of $m> 1 M_\odot$ (corresponding to a completeness of higher
than 60\%) were taken into consideration for the IMF computation, which is
indicated by the limits of the IMF line in Fig. \ref{n2194imfplot}.

Note that as a consequence of the age of NGC\,2194, the cluster may have
encountered dynamical evolution during its lifetime, which has to be kept in
mind when using the term ``{\it initial} mass function''. However, we follow
Scalo's (\cite{scalo1}) nomenclature, who uses the expression for intermediate
age clusters, to discriminate between a mass function based on the initial and
the present day stellar masses.

\section{Summary and discussion}

\label{caconcl}

With our work we found NGC\,1960 to be a young open star cluster with an age
of 16 Myr. It is located at a distance of 1300 pc from the Sun. These results
confirm the findings of Barkhatova et al. (\cite{barkhatova}) obtained with
photographic photometry.

We derived proper motions of 404 stars in the region of the cluster down to
$14 \mbox{\ mag}$. 178 of those can be considered members of NGC\,1960. Despite
the problems with our proper motion determination (see Sect. \ref{n1960eb}),
we are able to state that our results do not support the values given as the
absolute proper motion of NGC\,1960 by Glushkova et al. (\cite{glush}) on the
base of the ``Four Million Star Catalog'' (Gulyaev \& Nesterov \cite{4M}):
They found $\mu_\delta=-8.2 \pm 1 \mbox{\ mas yr}^{-1}$ which is in agreement
with our study, but $\mu_\alpha \cos \delta=14.7 \pm 1 \mbox{\ mas yr}^{-1}$
which differs from our result by more than $10 \mbox{\ mas yr}^{-1}$.

Our study of the IMF of NGC\,1960 led to a power law with a slope of
$\Gamma=-1.23 \pm 0.17$. This value is very high (i.e. the IMF is shallow)
compared to other studies, however, it still matches the interval for
$\Gamma$ suggested by Scalo (\cite{scalo2}) for intermediate mass stars ($-2.2
\leq \Gamma \leq -1.2$).

Although we should stress that we cannot say anything about the shape of the
IMF in the very low mass range ($m \ll M_\odot$), we do not see any evidence
for a flattening of the IMF of NGC 1960 below $1 M_\odot$.

NGC\,2194 --- with an age of 550 Myr --- belongs to the intermediate age
galactic open star clusters. Our findings from the photometric study are in
good agreement with the photographic $RGU$ photometry published by del\,Rio
(\cite{delrio}).

As the cluster is located at a distance of almost 3 kpc we could only cover
its mass spectrum down to $1 M_\odot$. Nevertheless, we were able to
determine the IMF on the base of 623 main sequence stars which led to a slope
of $\Gamma=-1.33 \pm 0.29$, almost Salpeter's (\cite{salpeter}) value, but
still close to the shallow end of the interval given by Scalo (\cite{scalo2}).

In our previous paper (Sanner et al. \cite{n0581paper}), we studied the open
star cluster NGC\,581 (M\,103) for which we found the same age of $16 \pm 4$
Myr as for NGC\,1960, but a much steeper IMF slope of $\Gamma=-1.80 \pm
0.19$. We therefore can state that our method of IMF determination does not
systematically lead to steep or shallow mass functions.

With our yet very small sample, it is not possible to find evidence for the
dependence of the IMF of open star clusters on any parameter of the
cluster. Therefore, we will have to investigate further clusters and also
compare our results with other studies.

\acknowledgements

The authors thank Wilhelm Seggewiss for allocating observing time at the
telescopes of Hoher List Observatory, Santi Cassisi for providing the
isochrones necessary for our study and Andrea Dieball and Klaas S.~de Boer for
carefully reading the manuscript. J.S. thanks Georg Drenkhahn for his valuable
hints concerning the maximum likelihood analysis software. M.A. and
J.B. acknowledge financial support from the Deutsche Forschungsgemeinschaft
under grants Bo 779/21 and ME 1350/3-2, respectively. This research has made
use of NASA's Astrophysics Data System Bibliographic Services, the CDS data
archive in Strasbourg, France, and J.-C. Mermilliod's WEBDA database of open
star clusters.

\end{document}